\documentclass[twocolumn]{aastex631}

\usepackage{rotating}

\begin{document}

\title{FAST H{\sc i} 21 cm study of blueberry galaxies}

\author[0000-0002-3698-3294]{Yogesh Chandola}
\email{yogesh.chandola@pmo.ac.cn}
\affiliation{Purple Mountain Observatory, Chinese Academy of Sciences,
10 Yuan Hua Road, Qixia District,
Nanjing 210023, People's Republic of  China}
\author[0000-0002-9390-9672]{Chao-Wei Tsai}
\email{cwtsai@nao.cas.cn}
\affiliation{National Astronomical Observatories, Chinese Academy of Sciences, 20A Datun Road, Beijing 100101,
People's Republic of China}
\affiliation{Institute for Frontiers in Astronomy and Astrophysics, Beijing Normal University, Beijing 102206, People's Republic of China}
\affiliation{School of Astronomy and Space Science, University of Chinese Academy of Sciences, Beijing 100049,
People's Republic of China}
\author[0000-0002-4464-8023]{D.J. Saikia}
\affiliation{Inter-University Centre for Astronomy and Astrophysics (IUCAA), Pune 411007, India}
\author[0000-0003-4007-5771]{Guodong Li}
\affiliation{Kavli Institute for Astronomy and Astrophysics (KIAA), Peking University, Beijing 100871, People's Republic of China}
\affiliation{National Astronomical Observatories, Chinese Academy of Sciences, 20A Datun Road, Beijing 100101,
People's Republic of China}
\author[0000-0003-3010-7661]{Di Li}
\affiliation{Department of Astronomy, Tsinghua University, Beijing 100084, People's Republic of China}
\affiliation{National Astronomical Observatories, Chinese Academy of Sciences, 20A Datun Road, Beijing 100101,
People's Republic of China}
\author[0000-0001-8108-0986]{Yin-Zhe Ma}
\affiliation{Department of Physics, Stellenbosch University, Matieland 7602, South Africa}



\begin{abstract}
Green Peas (GPs) and blueberry galaxies (BBs) are thought to be local analogs ($z<$0.1) of high redshift Ly$\alpha$ emitters. H{\sc i} study of these can help us understand the star formation in the primordial Universe. In this Letter, we present the results of H{\sc i} 21 cm study of 28 high specific star formation rate (sSFR $\gtrsim$10$^{-8}$ yr$^{-1}$) BBs at $z\lesssim$0.05 with the Five-hundred-meter Aperture  Spherical radio Telescope.  We report significant H{\sc i} detection towards two BBs namely J1026+0426 and J1132+0809, and discuss possible H{\sc i} contribution from neighboring galaxies. The median 3$\sigma$ upper limit of $\sim$2.0$\times$10$^{8}$ M$_{\odot}$ was obtained on H{\sc i} mass for galaxies with nondetections. We find BBs tend to have lower H{\sc i}-to-stellar mass ratio or gas fraction  ($f_{\rm HI}$) than expected from $f_{\rm HI}$-sSFR and $f_{\rm HI}$-$M_{\ast}$ relations for main-sequence galaxies. The BBs also have a median 3$\sigma$ upper limit on H{\sc i} gas depletion time scale ($\tau_{\rm HI}$) $\sim$0.5 Gyr, about 1 order of magnitude lower than $\tau_{\rm HI}$ for local main-sequence galaxies. We find a significantly low H{\sc i} detection rate of 2/28 (7.1$^{+9.4}_{-4.6}$ \%) towards these galaxies, which is similar to previous H{\sc i} studies of low redshift GPs of high ionization parameter indicator, O32 $\equiv$O[{\sc iii}]$\lambda$5007/O[{\sc ii}]$\lambda$3727  ratios $\gtrsim$10.

\end{abstract}
%which include the possible H{\sc i} from neighbouring galaxies}.
\keywords{Galaxies (573); H {\sc i} emission (690); Radio spectroscopy(1359); Starburst galaxies(1570); Blue compact dwarf galaxies(165)}


\section{Introduction} \label{sec:intro}
%\subsection{Blueberries: introduction} 
To understand star-formation near reionization in the early primordial Universe, there have been many searches for high redshift galaxies, such as the Ly$\alpha$ emitters (LAE; \citealt{gawiser2007ApJ...671..278G}). LAEs are compact, low stellar mass galaxies with very low metallicity and strong Ly$\alpha$ emission in the high redshift ($z >$ 2) Universe. However, because of the instrumental limitations, it is difficult to study high redshift objects comprehensively. Hence, the properties of local LAE analogs such as Green Peas (GPs; \citealt{cardamone2009MNRAS.399.1191C}) provide crucial hints about the physical processes in these compact and low metallicity systems. GPs like LAEs are compact, low metallicity systems at relatively lower redshifts ($z\lesssim$ 0.3). Studies by \cite{izotov2016MNRAS.461.3683I,izotov2018MNRAS.474.4514I,izotov2018MNRAS.478.4851I} find Lyman continuum (LyC) leakage in significant amounts (2\%-72\%) in some GP galaxies sufficient to ionize the neutral hydrogen in the intergalactic medium. Recently, \cite{yang2017ApJ...847...38Y} discovered a similar class of galaxies termed \lq blueberry' galaxies (BBs) in the nearby Universe ($z \lesssim$ 0.05). These galaxies are very compact ($\lesssim$ 1 kpc in linear projected size) and have small a star formation rate (SFR), but these are high specific star formation rate (sSFR $>$ 10$^{-8}$ yr$^{-1}$) starburst systems. These systems are characterized by their extreme emission line ratios (O[{\sc{iii}}]$\lambda$5007/O[{\sc{ii}}]$\lambda$3727 $\sim$8-60) and extreme blue colors ($g-r$ $< -$0.5 mag and $r-i$ $< $ 1 mag). Most of these have very low stellar mass ($M_{\ast}$ $\sim$ 10$^{6.5-8.5}$ $M_{\odot}$) and some of these systems also have very low metallicities ($Z$ $\lesssim$1/10 $Z_{\odot}$) which make them similar to higher redshift ($z >$ 0.1) GPs and LAEs. GPs owe their green color to the strong O[{\sc iii}]$\lambda$5007 emission line with large equivalent widths ($\sim$1000 \AA) falling in the Sloan Digital Sky Survey \citep[SDSS;][]{abazajian2009ApJS..182..543A} $\lq r$' band while for lower redshift BBs it falls in the $\lq g$' band making their color appear blue. Compared to several other star-forming dwarf galaxies like local blue compact dwarfs \citep[BCDs;][]{gildepaz2003ApJS..147...29G}, these systems have similar stellar mass, but higher emission line strength and gas ionization, and hence reside at extreme top-left in the Baldwin-Phillips-Terlevich \citep[BPT;][]{bpt1981PASP...93....5B} diagram.
%%%%%%%%%%%%%%%%%

H{\sc i} 21cm line can be used to trace the cold gas fuel reservoir for these systems.  Though there have been earlier H{\sc i} studies on the starburst BCDs, most of them have sSFRs $<$ 10$^{-8}$ yr$^{-1}$ \citep{chandola2024MNRAS.527..603C}. In our recent work, we studied H{\sc i} gas contents toward 11 mid-infrared (IR) bright BCDs of sSFR $\sim$ 10$^{-8}$ yr$^{-1}$ with the Arecibo telescope, where we have detections toward six sources \citep{chandola2024MNRAS.527..603C}.  Mid-IR bright BCDs observed with the Arecibo telescope were found to have a very low H{\sc i} gas depletion time scale of $\sim$0.3 Gyr \citep{chandola2024MNRAS.527..603C}. It indicates that these systems will deplete their gas reservoir quite fast due to their high SFR. Previously, from a sample of 40 GPs at low redshift $z<$0.1 selected from \cite{jiang2019ApJ...872..145J}, \cite{kanekar2021ApJ...913L..15K} found that GPs have a very low depletion time scale of $\sim$0.6 Gyr and sources with higher O[{\sc iii}]$\lambda$5007/O[{\sc ii}]$\lambda$3727 have low H{\sc i} detection rates. Like BCDs in our Arecibo study \citep{chandola2024MNRAS.527..603C} and GPs from \cite{kanekar2021ApJ...913L..15K}, BBs are very high sSFR systems but have lower SFR and slightly lower stellar mass.  Hence, to get a complete picture, it is important to probe whether these systems have similar low depletion time scales for different ranges of stellar masses and SFRs. BBs also have very high O[{\sc{iii}}]$\lambda$5007/O[{\sc{ii}}]$\lambda$3727 (O32) ratios and hence form a useful sample to test the presence of an atomic H{\sc i} reservoir in extreme conditions. Recently, \cite{dutta2024MNRAS.531.5140D} studied H{\sc i} content toward one of the BBs, J1509+3731, to find a very low depletion time scale of $\sim$0.2 Gyr indicating fast depletion of H{\sc i} reservoir. Interferometric H{\sc i} studies of BBs and nearby GPs suggest a merger origin \citep{purkayastha2022ApJ...933L..11P,dutta2024MNRAS.531.5140D}.
%%%%%%%%%%%%%%%
\begin {table*}
\begin {center}
{\scriptsize
\caption {Characteristics of observed BBs}     
	\begin {tabular}{l r l c c c c c c c c c}
	\hline
 \hline
	(1)   & (2)& (3)       & (4)          & (5) & (6) & (7) & (8) & (9) & (10)&(11)&(12) \\ 
	Source Name (ID)& $D_{\rm L}$&Date of Ob. & Beam(s) Used & Time   &$\Delta$$S_{rms}$   & log $M_\mathrm{HI}$ & log $M_\mathrm{\ast}$& log SFR & log $f_\mathrm{ HI}$ & log $\tau_\mathrm{HI}$ & log O32\\
	&  (Mpc) & & & (s) &    (mJy)  & ($M_{\odot}$)     & ($M_{\odot}$)   & ($M_{\odot}$ yr$^{-1}$)& &(yr) &     \\
	\hline
J0146+0319 (27) & 207.1 & 2022 Aug 22& M01,M08 &360   &  0.32 & $<$8.34 & 7.7 & 0.06 & $<$0.64 &$<$8.27 & 1.07\\
J0216+1715 (28) & 173.0 & 2022 Aug 22& M01,M08 &1980  &  0.15 & $<$7.85 & 6.8 & $-$0.67 & $<$1.05 & $<$8.53 & 0.99\\
J0238+0124 (31) & 221.4 & 2022 Aug 21& M01,M08 & 720  &  0.28 &$<$8.34 & 7.5 & $-$0.36 & $<$0.84 & $<$8.70 & 0.92 \\
J0357+1808 (34) & 164.4 & 2022 Aug 24& M01,M08 & 180  &  0.63 &$<$8.43 & 8.3 & $-$0.27 & $<$0.13 & $<$8.70 & 1.36  \\
J0820+5431 (2) & 170.1  & 2023 Apr 12 & M01\tablenotemark{\scriptsize{a}} & 3240    &  0.19 &$<$7.94 & 6.6 & $-$0.82 & $<$1.34 & $<$8.76 & 1.33  \\
J0825+1846 (3) & 167.1  & 2022 Oct 7 & M01,M08& 540  &  0.41 & $<$8.26 & 7.2 & $-$0.30 & $<$1.06 & $<$8.54 & 1.08\\
J0827+1059 (39) & 192.7  & 2022 Oct 7 & M01,M08& 900  &  0.25 &$<$8.17 & 7.2 & $-$0.46 & $<$0.97 & $<$8.62 & 1.17\\
J0837+1823 (41) & 180.8 & 2022 Oct 8& M01,M08& 540   &  0.33 & $<$8.23 & 7.3 & $-$0.31 & $<$0.93 & $<$8.54 & 1.61\\
J0922+6324 (47) & 173.4 & 2022 Oct 26 & M01,M08& 1440 &  0.37 &$<$8.25 & 6.9 & $-$0.71 & $<$1.35 &$<$8.96 & 0.98 \\
J0926+4504 (4)& 186.7  & 2022 Oct 8 & M01\tablenotemark{\scriptsize{a}} & 900    &  0.33 &$<$8.26 & 7.2 & $-$0.64 &$<$1.06 & $<$8.90 & 1.46  \\

J1026+0426 (54)& 186.3 & 2021 Feb 15& M01\tablenotemark{\scriptsize{a}} & 2160     &  0.16 &7.73\tablenotemark{\scriptsize{b}} & 7.1 & $-$0.66 &0.63 & 8.39 & 0.79\\

J1032+4919 (5)& 194.8  & 2022 Oct 8 & M01\tablenotemark{\scriptsize{a}} & 360     &  1.24 &$<$8.87 & 8.0 & $-$0.05 & $<$0.87 & $<$8.93 & 1.32  \\
J1035+1400 (57) & 175.2 & 2022 Oct 23 & M01,M08 &180  &  0.79 &$<$8.58 & 7.6 & $-$0.85 & $<$0.98 & $<$9.43 & 1.10\\
J1113+0301 (66) & 101.8  & 2022 Oct 23 & M01,M08 &360  &  0.49 &$<$7.91 & 6.8 & $-$0.66 & $<$1.11 & $<$8.57 & 1.29 \\
J1123+2050 (67) & 144.0  & 2022 Oct 23 & M01,M08& 180  &  0.82 &$<$8.43 & 8.6 & 0.17 &$<-$0.17 & $<$8.26 & 0.97 \\
J1132+0809 (68) & 219.6 & 2022 Nov 7 & M01,M08 & 1620&  0.17 & 8.24\tablenotemark{\scriptsize{c}}& 7.2 & $-$0.24 & 1.04 & 8.48 & 1.27\\
 ..& .. & .. & .. &.. & .. & 9.56\tablenotemark{\scriptsize{d}}& 7.2 & $-$0.24 & 2.36 & 9.80 & 1.27\\
J1136+3427 (69) & 153.5 & 2022 Nov 4 & M01,M08& 1620 &  0.22 &$<$7.91 & 6.7 & $-$1.31 & $<$1.21 & $<$9.22 & ..  \\
J1139+0040 (70) & 184.0 & 2022 Nov 17 & M01,M08 & 180 &  0.80 &$<$8.63 & 8.0 & $-$0.47 &$<$ 0.63 & $<$9.11 & 1.05 \\
J1323$-$0132 (6)& 97.8   & 2022 Dec 18 & M01,M08& 180  &  0.90 & $<$8.13 & 7.1 & $-$0.43 &$<$ 1.03 & $<$8.56 & .. \\
J1347+0755 (88) & 193.3 & 2022 Dec 13 &M01,M08& 180   &  1.05 &$<$8.79 & 8.5 & $-$0.27 &$<$ 0.29 & $<$9.06 & 1.04\\
J1355+4651 (7) & 122.9   & 2022 Dec 18 &M01\tablenotemark{\scriptsize{a}} & 540      &  0.50 &$<$8.08 & 6.8 & $-$0.66 &$<$ 1.28 & $<$8.74 & 1.18  \\
J1400+1951 (90) & 237.2 & 2023 Apr 12 &M01,M08 & 2880 &  0.17 &$<$ 8.18 & 7.1 & $-$0.63 &$<$ 1.08 &$<$ 8.81 & 1.33\\
J1444+0409 (9) & 170.9   & 2022 Dec 24 &M01,M08 &540   &  0.48 &$<$8.35 & 7.3 & $-$0.51 & $<$1.05 & $<$8.85 & 1.20 \\
J1509+3731 (10) & 143.0  & 2022 Aug 24 &M01\tablenotemark{\scriptsize{a}} & 180      &  1.11 & $<$8.56 & 8.1 & 0.21 &$<$ 0.46 &$<$ 8.35 & 1.20\\
J1556+4806 (12) & 223.3 & 2022 Aug 24 & M01\tablenotemark{\scriptsize{a}} & 360     &  0.50 &$<$8.60 & 7.9 & 0.03 &$<$ 0.70 &$<$ 8.56 & 0.98 \\
J1602+1445 (106) & 159.9 & 2023 Jan 11 & M01,M08 & 180 &  1.10 &$<$8.65 & 8.0 & $-$0.44 &$<$ 0.65 &$<$ 9.09 & 1.22 \\
J1608+3528 (13) & 143.6  & 2023 Jan 4 & M01\tablenotemark{\scriptsize{a}} & 180     &  1.29 &$<$ 8.62 & 7.5 & $-$0.35 & $<$1.12 & $<$8.97 & 1.55\\
J2320+1225 (122) &185.2  & 2023 May 18 & M01,M08 & 2700&  0.15 &$<$ 7.91 & 6.8 & $-$0.69 &$<$ 1.11 &$<$ 8.60 & 1.00 \\

	\hline                          
	\end {tabular}
 \label{sourchar1}  
		\tablecomments{
		Column (1): Source name in increasing R.A. order with the identifiers used by \cite{yang2017ApJ...847...38Y} in brackets.
		Column (2): $D_{\rm L}$, luminosity distance in Mpc, estimated using cosmological parameters, $H_{\rm o}=$70 km s$^{-1}$ Mpc$^{-1}$, $\Omega_{m}=$0.3, $\Omega_{\Lambda}=0.7$.
   Column (3): date of observation.
  Column (4): beam(s) used in final H{\sc i} spectra.
		 Column (5): ON target source time per beam in seconds.
		Column (6): $\Delta S_{\rm rms}$ (1$\sigma$), noise level per $\sim$10 km s$^{-1}$ in mJy.
		Column (7): log $M_{\rm HI}$, logarithm of the total H\,{\sc{i}} mass in $M_{\odot}$. Column (8): log $M_{\ast}$, logarithm of the total stellar mass in $M_{\odot}$ from \cite{yang2017ApJ...847...38Y}. Column (9): logarithm of star formation rate in $M_{\odot}$ yr$^{-1}$ from \cite{yang2017ApJ...847...38Y}. Column (10): logarithm of H{\sc i} gas fraction ($f_{\rm HI}$). Column (11): logarithm of H{\sc i} gas depletion time scale in years. Column (12): logarithm of O32$\equiv$O[{\sc iii}]$\lambda$5007/O[{\sc ii}]$\lambda$3727 ratio estimated from the line fluxes in the ViZier Online data catalog (\url{https://cdsarc.cds.unistra.fr/viz-bin/cat/J/ApJ/847/38}) of \cite{yang2017ApJ...847...38Y}.} 
  
   \tablenotetext{\scriptsize{a}}{The coordinates for beam M08 do not coincide with the source position or ON source coordinates for beam M01. Hence only the spectrum from beam M01 has been used.}
  \tablenotetext{\scriptsize{b}}{H{\sc i} mass of BB J1026+0426 calculated assuming that the ratio of H{\sc i} mass of BB to H{\sc i} mass of the neighboring galaxy is equal to their stellar mass ratio. The total H{\sc i} mass is estimated to be 10$^{9.84}$ $M_{\odot}$ using integrated line flux density of 0.84$\pm$0.05 Jy km s$^{-1}$. The spectral line for J1026+0426 has H{\sc i} velocity ($V_{\rm H {\sc{I}}}$) of 12662.3$\pm$2.2 km s$^{-1}$, FWHM of 148.8$\pm$4.4 km s$^{-1}$ and peak flux density of 6.9$\pm$0.4 mJy.}
  \tablenotetext{\scriptsize{c}}{H{\sc i} mass of BB J1132+0809 calculated assuming that the ratio of H{\sc i} mass of BB to H{\sc i} mass of the neighboring galaxy is equal to their stellar mass ratio. The total H{\sc i} mass is estimated to be 10$^{9.56}$ $M_{\odot}$ using integrated line flux density of 0.32$\pm$0.03 Jy km s$^{-1}$. The spectral line for J1132+0809 has H{\sc i} velocity ($V_{\rm H {\sc{I}}}$) of 14791.1$\pm$3.6 km s$^{-1}$, FWHM of 124.4$\pm$7.3 km s$^{-1}$ and peak flux density of 3.0$\pm$0.2 mJy.}
  \tablenotetext{\scriptsize{d}}{H{\sc i} mass of BB J1132+0809 calculated assuming that there is no neighboring galaxy at its redshift. It is equal to the total H{\sc i} mass $\sim$ 10$^{9.56}$ $M_{\odot}$ using integrated line flux density of 0.32$\pm$0.03 Jy km s$^{-1}$.}}

\end {center} 
\end {table*}
%%%%%%%%%%%%%%%%%%%%%%%%%%%%%%%%

In this Letter, we present the H{\sc i} observation results of a sample of 28 BBs from \cite{yang2017ApJ...847...38Y} with the Five-hundred-meter Aperture  Spherical radio Telescope \citep[FAST;][]{Nan2011IJMPD..20..989N} obtained during 2020-2023 (see Table.~\ref{sourchar1}). This is the largest sample of BBs with high O32 ratios studied in H{\sc i}. We describe the observations and data reduction process in Section~\ref{sec:obsdr}. We present and discuss results in  Section~\ref{results}. We summarize them in Section~\ref{sec:summary}. We use the concordance cosmology $H_{\rm o}=$70 km s$^{-1}$ Mpc$^{-1}$, $\Omega_{m}=$0.3, $\Omega_{\Lambda}=0.7$.
%%%%%%%%%%%%%%%%%%%%%%%%%%%%%%%%%%%%%%%%%%%%%%%%%%%%%%
\begin{figure}
    \centering
    \includegraphics[scale=0.36]{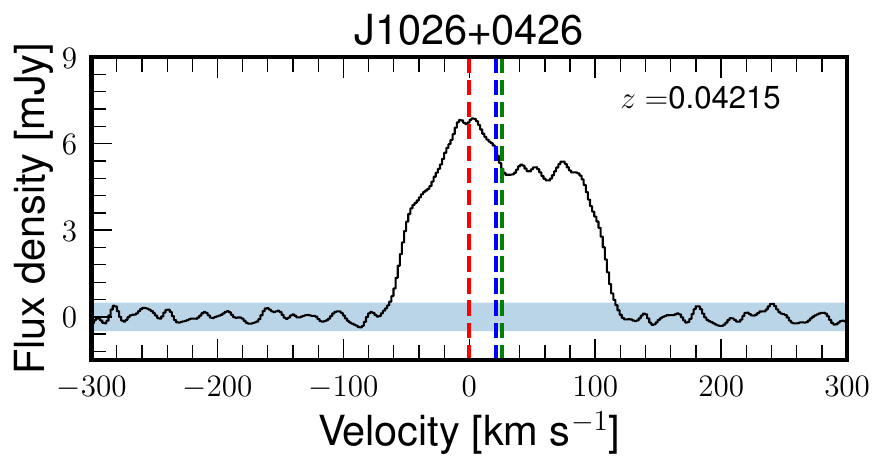}
    \includegraphics[scale=0.36]{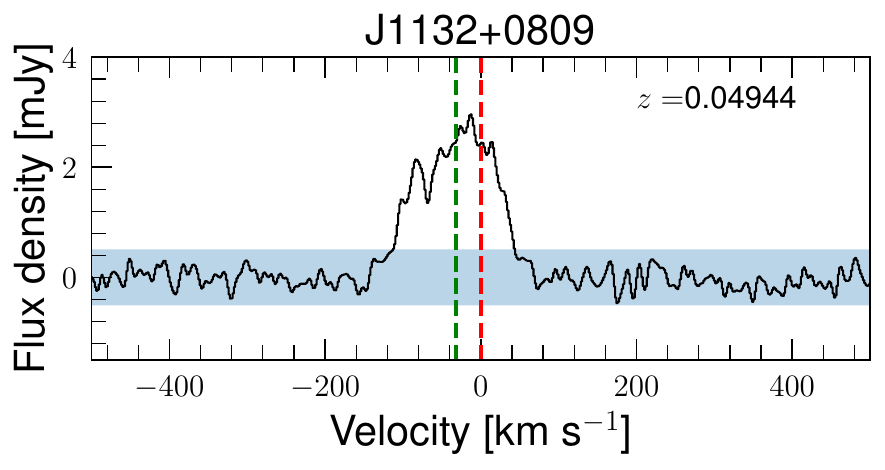}
  
    \caption{21 cm emission spectra of blueberry galaxies with H{\sc i} detections. The X-axes show the velocity relative to optical systemic velocity (cz) at the heliocentric frame in km s$^{-1}$. The Y-axes show the flux density in mJy. Blue shades mark the 3$\sigma$ noise levels. Red vertical dashed lines mark the zero relative velocity corresponding to the optical redshift of the blueberry galaxies. Green vertical dashed lines mark the H{\sc i} velocity ($V_{\rm HI}$) relative to the optical systemic velocity. In the upper panel, the blue dashed vertical line marks the optical systemic velocity of the neighboring spiral galaxy relative to the optical velocity of blueberry galaxy J1026+0426.}
    \label{fig:hispectra1}
\end{figure}

%%%%%%%%%%%%%%%%%%%%%%%%%%%%%%%%%%%%%%%%%%%%%%%%%%%%%

%%%%%%%%%%%%%%%%%%%%%%%%%%%%%%%%%%%%%%%%%%%%%%%%%%%%%%%%%%%%
\begin{figure*}
    \centering
    \includegraphics[scale=0.33]{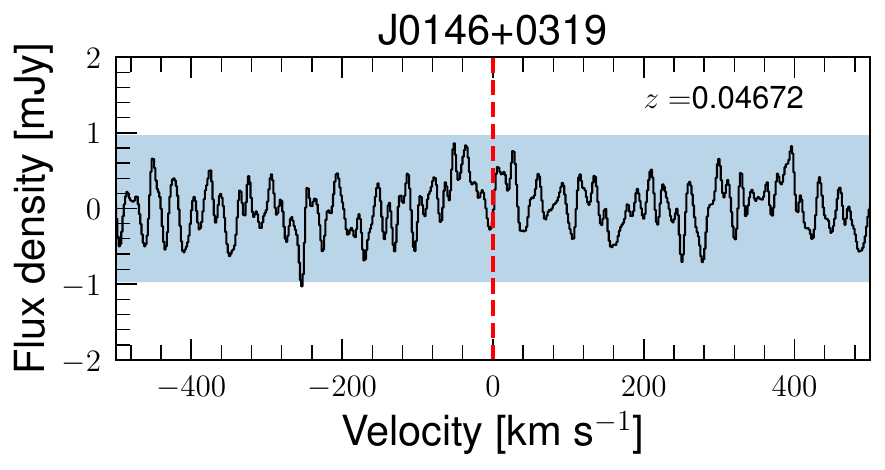}
    \includegraphics[scale=0.33]{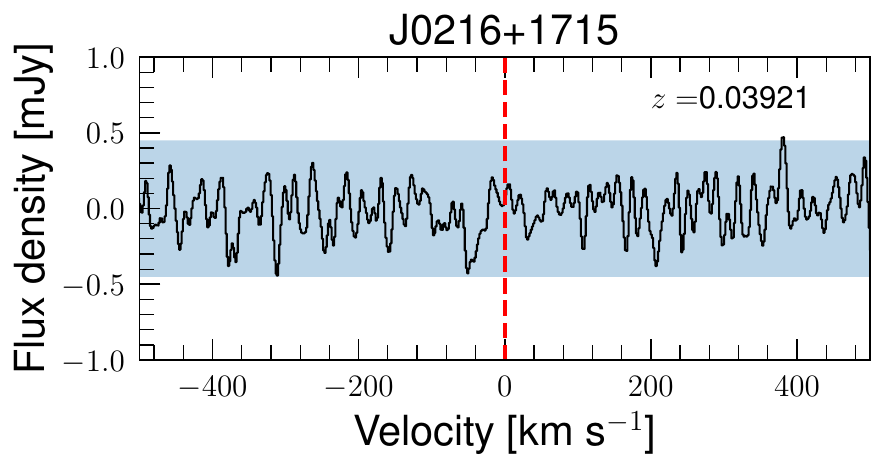}
    \includegraphics[scale=0.33]{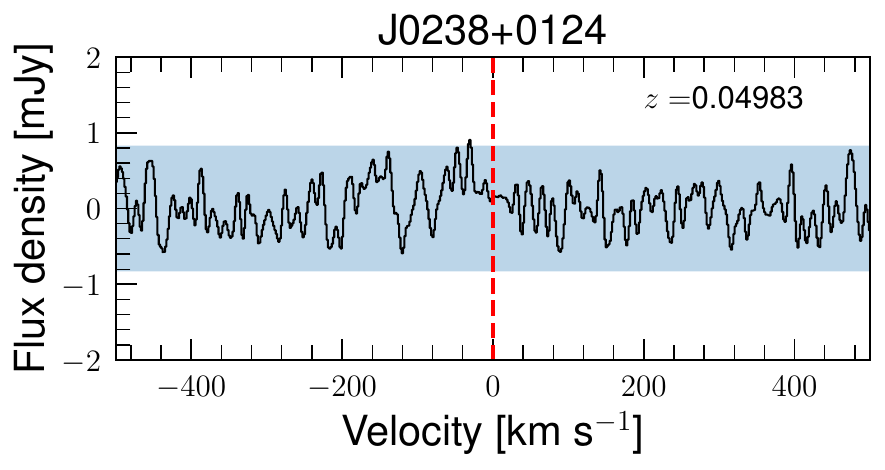}
    \includegraphics[scale=0.33]{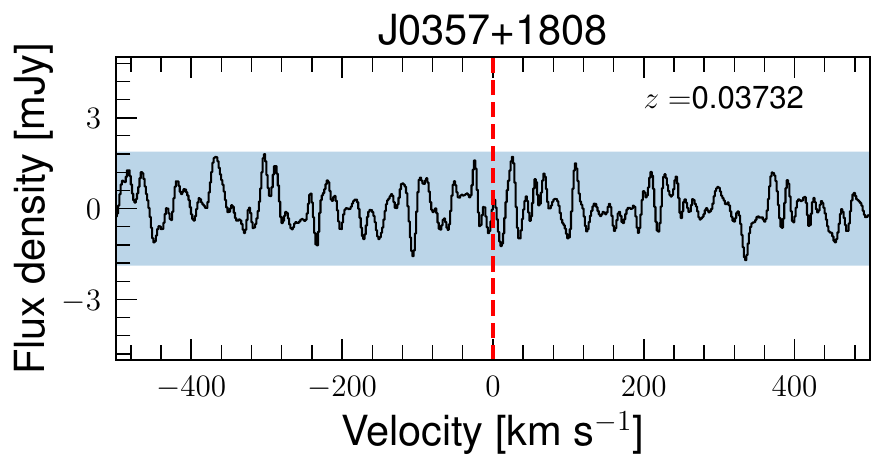}
    \includegraphics[scale=0.33]{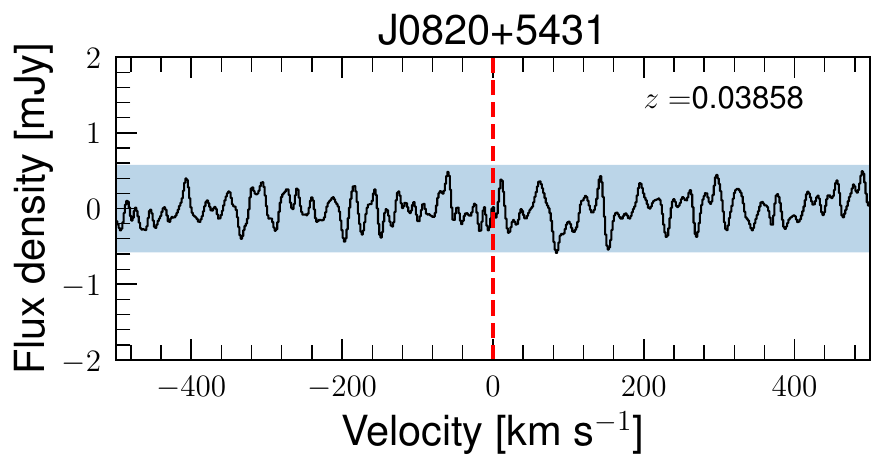}
    \includegraphics[scale=0.33]{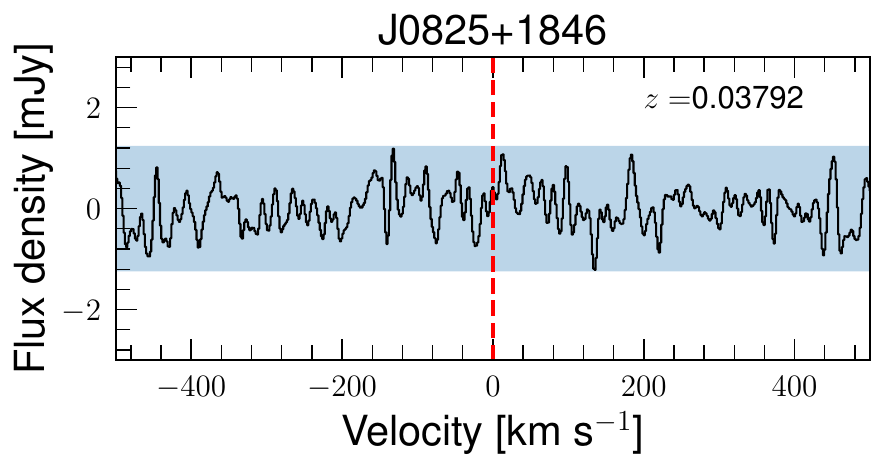}
    \includegraphics[scale=0.33]{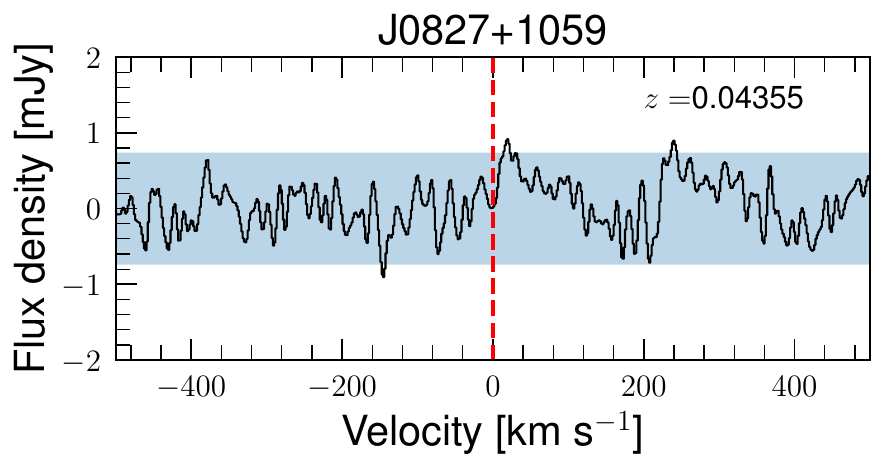}
    \includegraphics[scale=0.33]{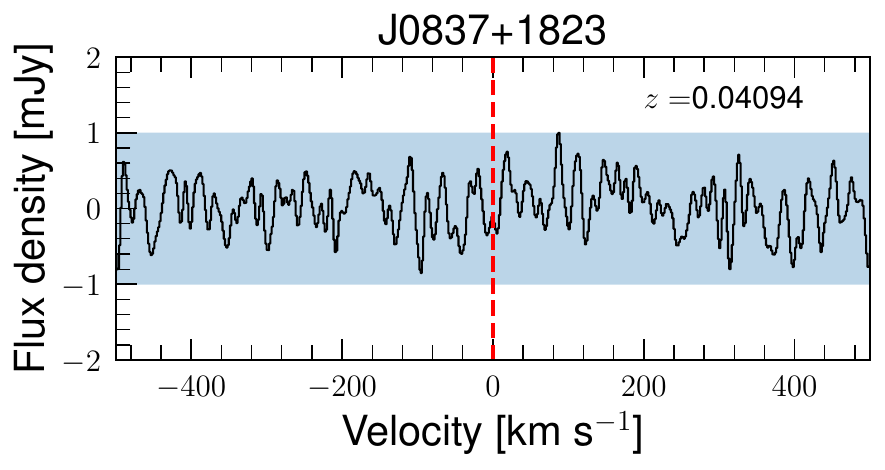}
    \includegraphics[scale=0.33]{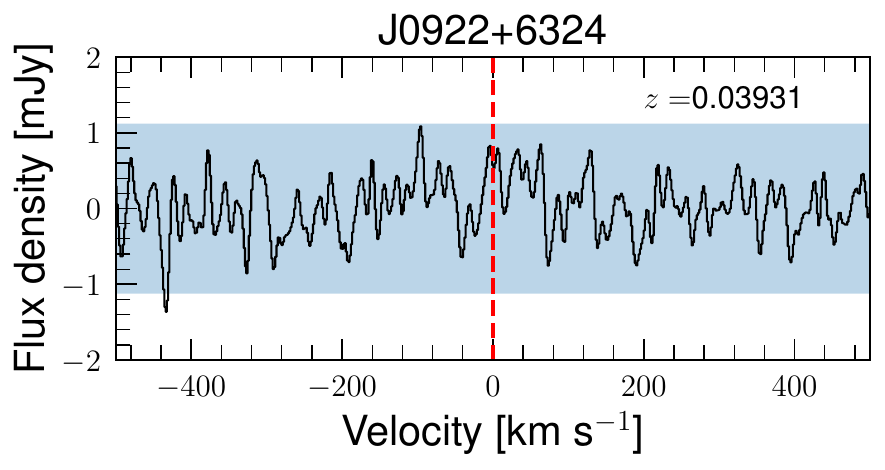}
    \includegraphics[scale=0.33]{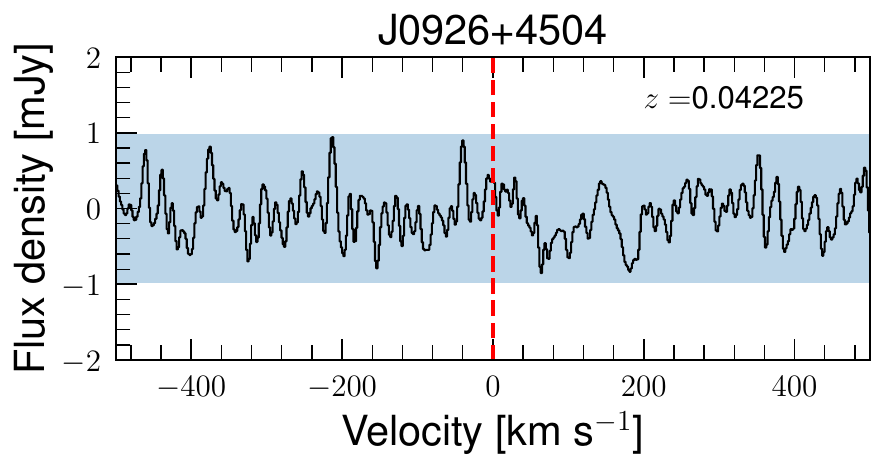}
    \includegraphics[scale=0.33]{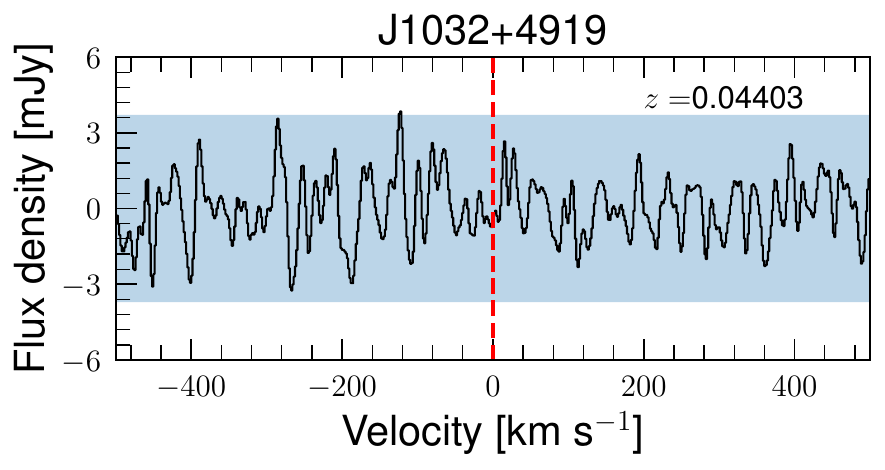}
    \includegraphics[scale=0.33]{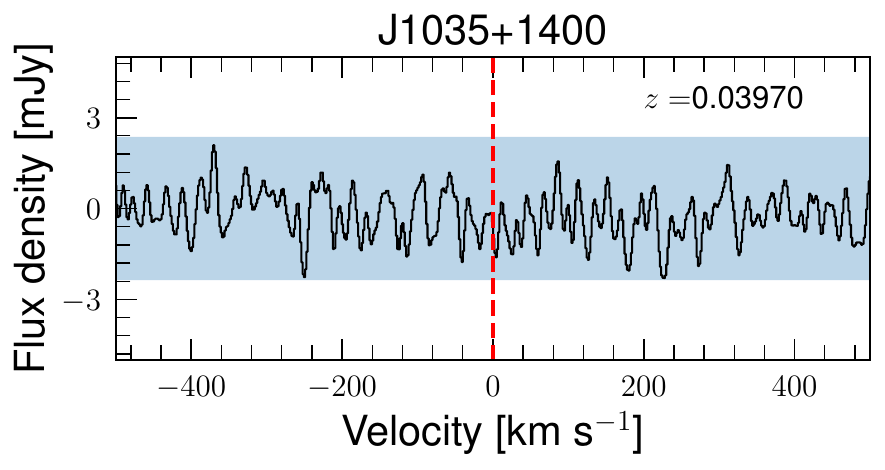}
    \includegraphics[scale=0.33]{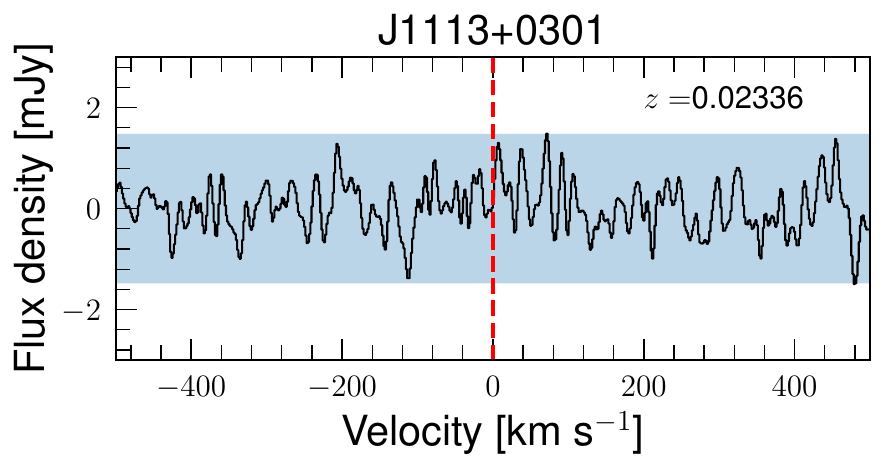}
    \includegraphics[scale=0.33]{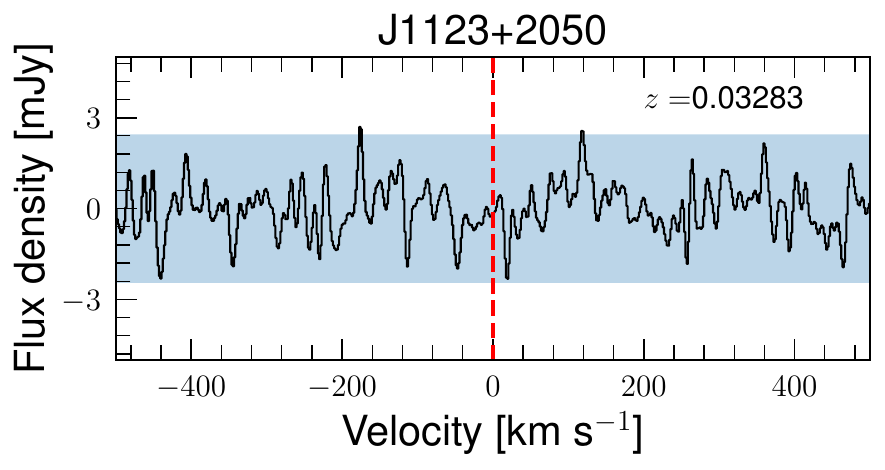}
    \includegraphics[scale=0.33]{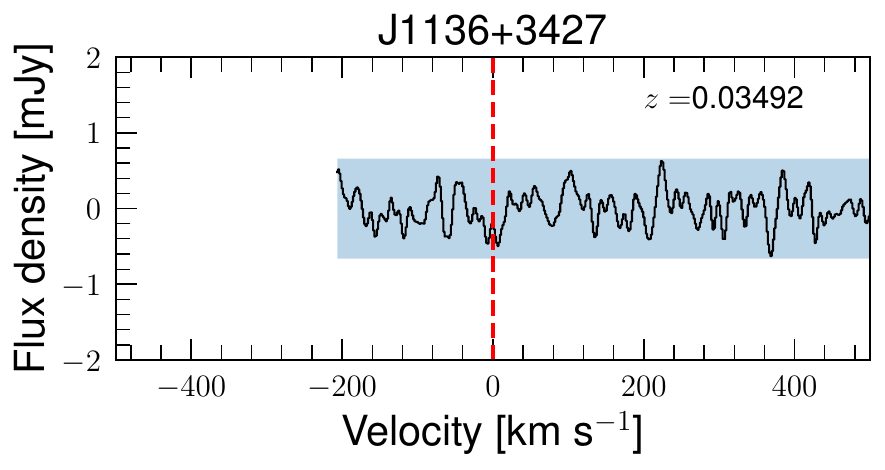}
     \includegraphics[scale=0.33]{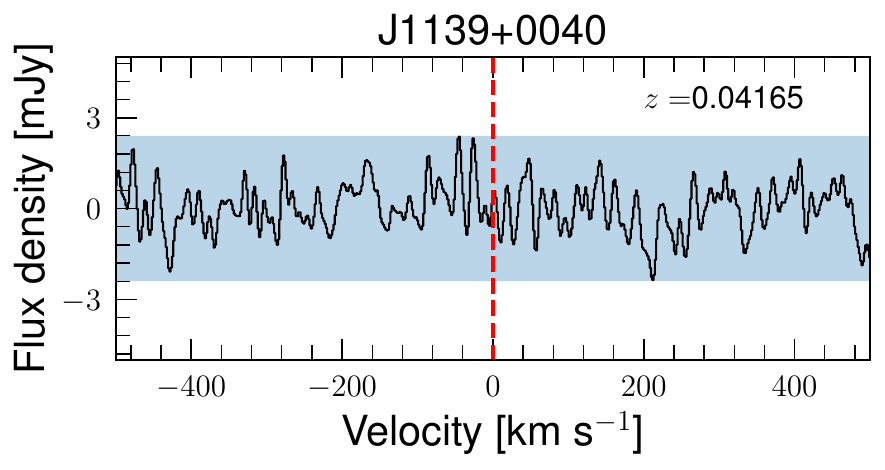}
    \includegraphics[scale=0.33]{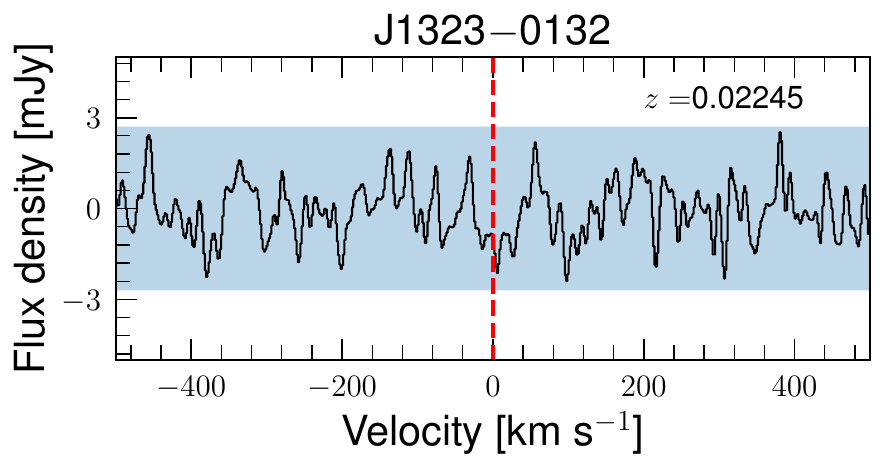}
    \includegraphics[scale=0.33]{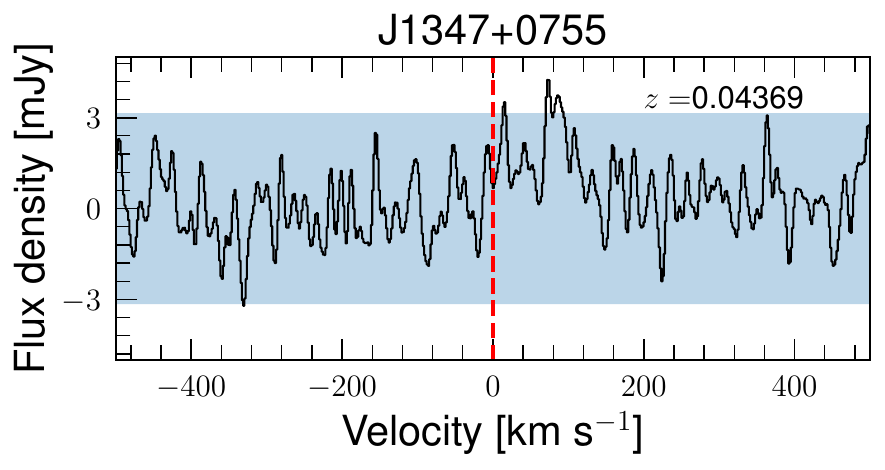}
    \includegraphics[scale=0.33]{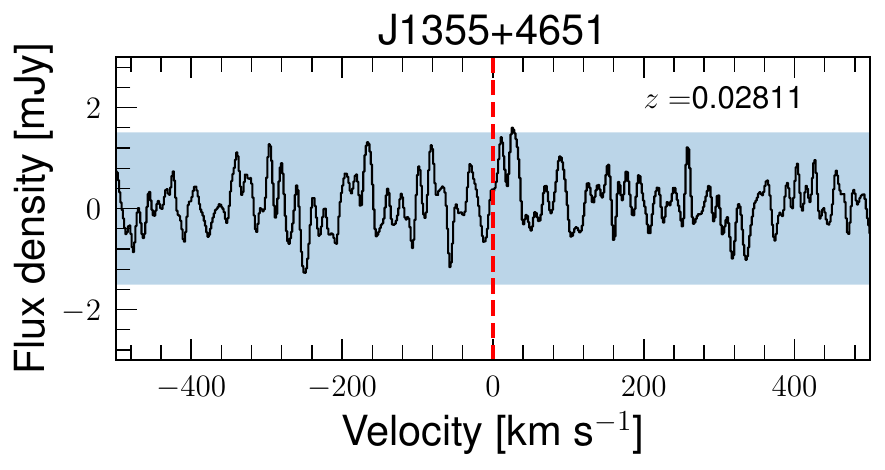}
    \includegraphics[scale=0.33]{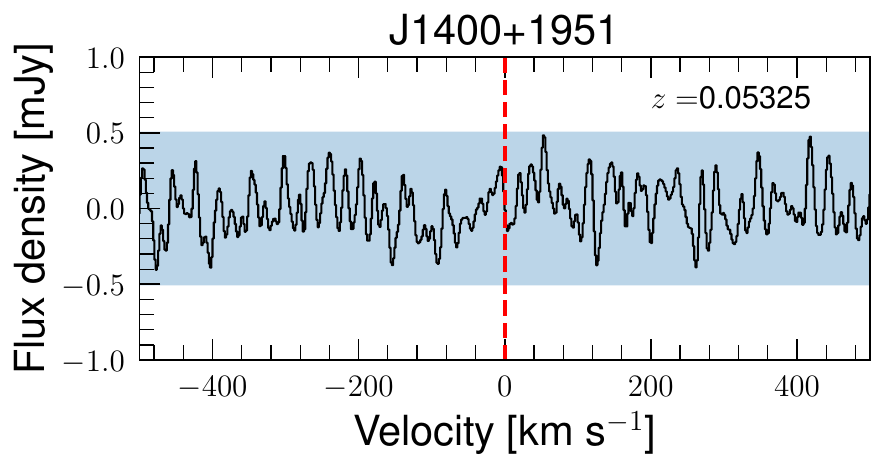}
    \includegraphics[scale=0.33]{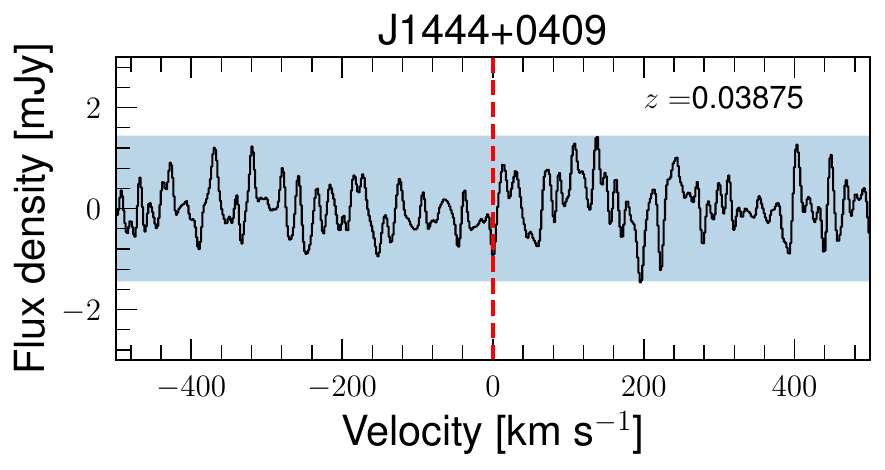}
    \includegraphics[scale=0.33]{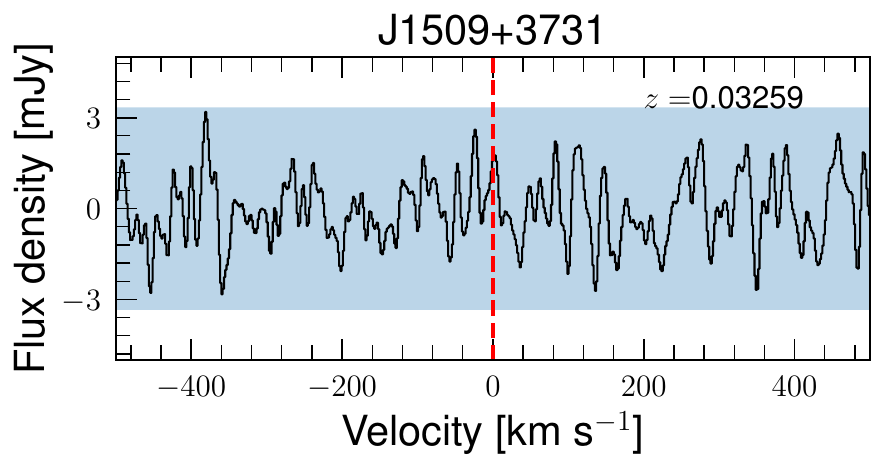}
     \includegraphics[scale=0.33]{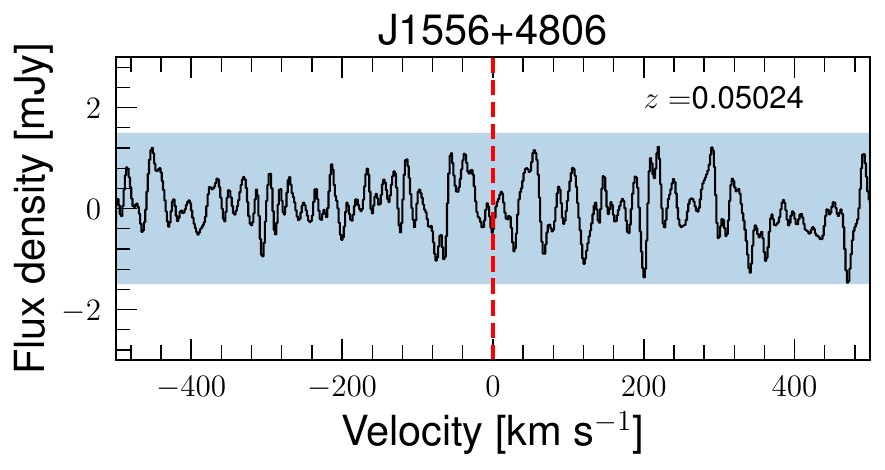}
    \includegraphics[scale=0.33]{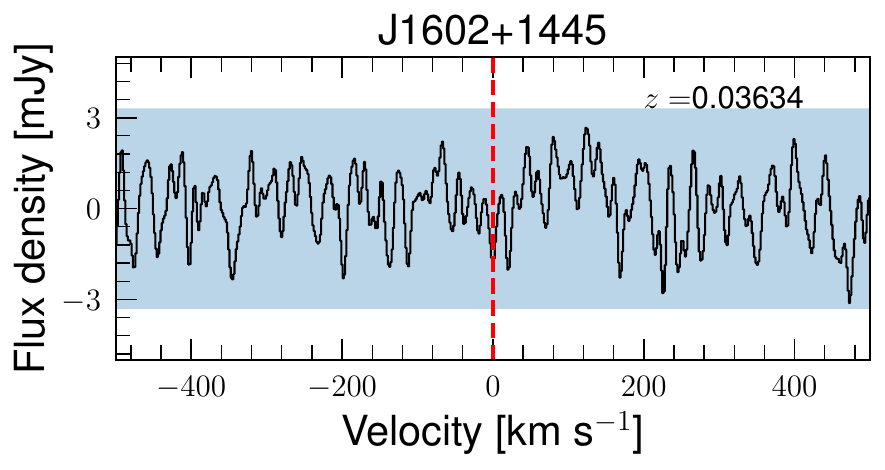} 
    \includegraphics[scale=0.33]{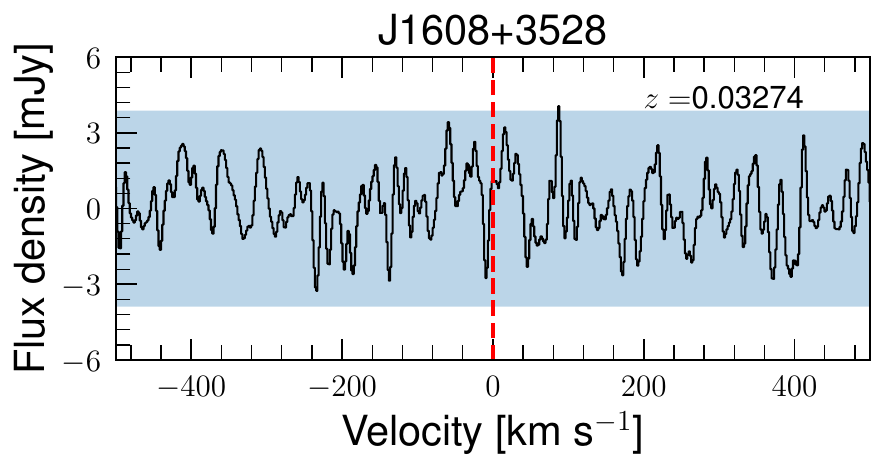}
     \includegraphics[scale=0.33]{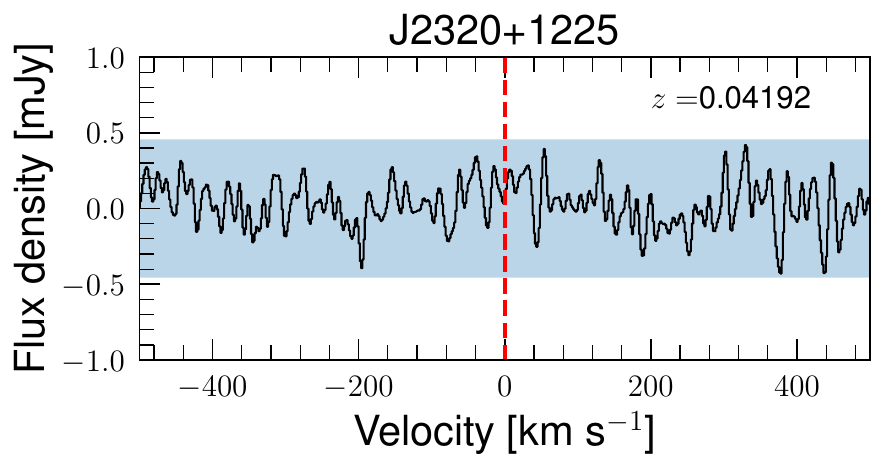}
    \caption{Spectra of blueberry galaxies with H{\sc i} nondetections. The symbols mean the same as in Fig.~\ref{fig:hispectra1}.}
    \label{fig:hispectra2}
\end{figure*}

%%%%%%%%%%%%%%%%%%%%%%
\section{Observations and data reduction}
\label{sec:obsdr}
The observations toward the BBs were executed in two phases. 
In the first phase from 2020 September to 2021 February, observations were carried out toward 21 sources for 15.1 hr (including overheads) using the standard ON and OFF position mode of FAST (project code: PT2020\_0082).  In this phase, the OFF position was made by shifting 18$\arcmin$ in the R.A. toward the west. However, the observations of 20 sources were severely affected due to the radio frequency interference (RFI) from the feed compressor. In the second phase 2022 August-2023 May, the observations were done for 27 sources for 19.2 hr (project code: PT2022\_0192), including those 20 sources that were affected by RFI in the first phase of observations. In the second phase of observations, the OFF position was made by shifting 11.$\arcmin$8 toward the west, so that we have another beam, M08, at the ON position of the target source while the central beam is at the OFF position.  We used a sampling time of 1 s and injected low (1 K) noise for 1 s after every 2 s in all the observations for calibration. The observational details along with source properties are provided in Table~\ref{sourchar1}. The total observing ON source time varies from 3 minutes to 96 minutes.

To reduce the data we used a self-developed Python-based pipeline. The pipeline first combines raw data and then separates the data from each ON-OFF cycle for processing. For each ON-OFF cycle, after separating the power values, $P^{\rm cal}_{\rm on}$ and $P^{\rm cal}_{\rm off}$ with calibration ON and calibration OFF, we use equations
\begin{equation}
    T^{\rm cal}_{\rm a,on} = P^{\rm cal}_{\rm on} T_{\rm cal}/(P^{\rm cal}_{\rm on}-P^{\rm cal}_{\rm off}) -T_{\rm cal},
\end{equation}
\begin{equation}
    T^{\rm cal}_{\rm a, off} = P^{\rm cal}_{\rm off} T_{\rm cal}/(P^{\rm cal}_{\rm on}-P^{\rm cal}_{\rm off})
\end{equation}
to convert power into antenna temperatures, $T^{\rm cal}_{\rm a, on}$ and $T^{\rm cal}_{\rm a, off}$ for calibration ON and OFF data, respectively. $T_{\rm cal}$ is the temperature of the noise injected for calibration. We then get the average of $T^{\rm cal}_{\rm a, on}$ and $T^{\rm cal}_{\rm a, off}$ to obtain antenna temperature, $T_{\rm a}$. We mask antenna temperature for RFI in time and frequency. The antenna temperature values are separated for ON and OFF positions, averaged in time, and then OFF position values are subtracted from ON position values to obtain brightness temperature,
\begin{equation}
    T_{\rm b}=T_{\rm a, ON} - T_{\rm a, OFF},
\end{equation}
where $T_{\rm a, ON}$ and $T_{\rm a, OFF}$ are antenna temperatures at the ON and OFF positions, respectively.
The brightness temperature values from all ON-OFF cycles are then averaged together. Thereafter, a sine function and high-degree polynomial are subtracted from two polarizations XX and YY separately to remove baseline ripples and continuum. After that, we average the data from two polarizations and convert K into Jy, using the gain values from \cite{jiang2020RAA....20...64J} for zenith angle $<$ 26.$^{\circ}$4 and equation (3) of \cite{zhang2019SCPMA..6259506Z} for zenith angle $>$ 26.$^{\circ}$4. The frequencies are converted to velocities and corrected to a heliocentric frame of rest using the \texttt{astropy} library. We averaged the spectra from two different beams M01 and M08 if their ON and OFF positions coincide on target sources. Finally, the spectra were Hanning smoothed to the velocity resolution of $\sim$10 km s$^{-1}$ from the original resolution of $\sim$7.63 kHz ($\sim$1.7 km s$^{-1}$).
%%%%%%%%%%%%%%%
\begin{figure*}
    \centering
    \includegraphics[scale=0.5]{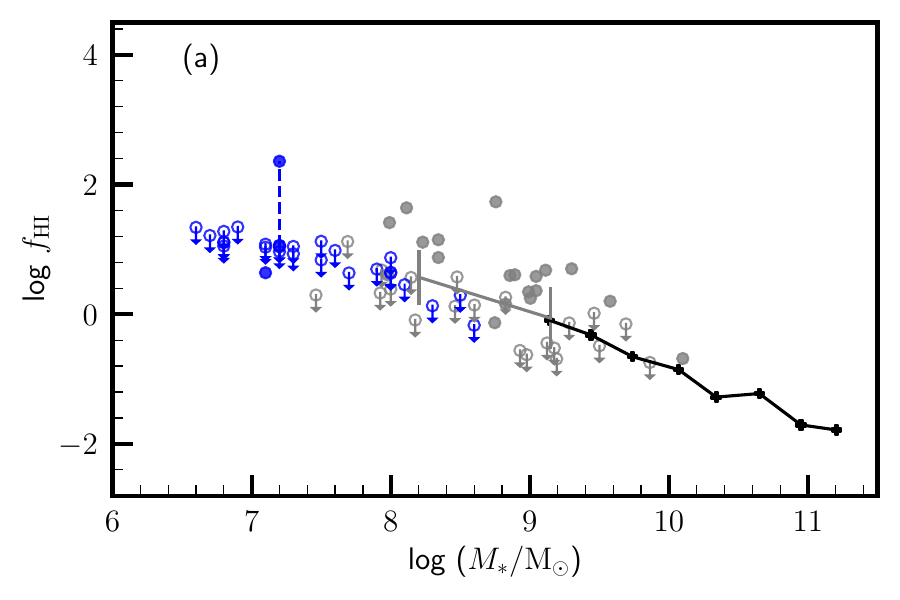}
    \includegraphics[scale=0.5]{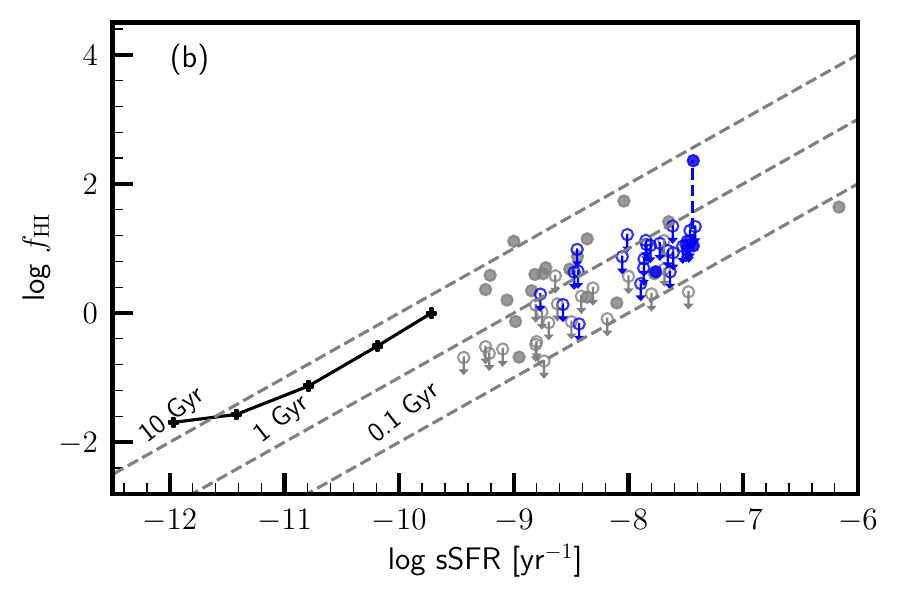}
     \includegraphics[scale=0.5]{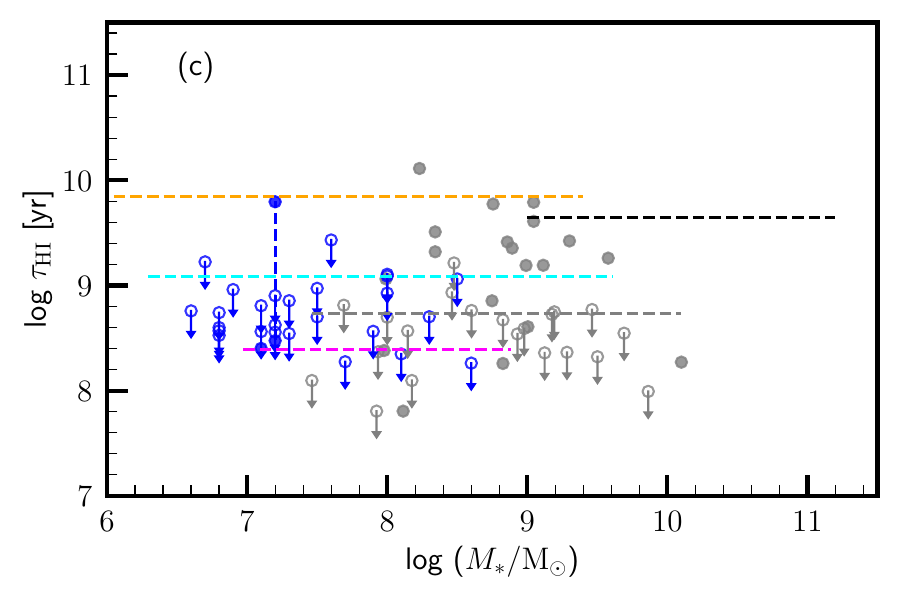}
      \includegraphics[scale=0.5]{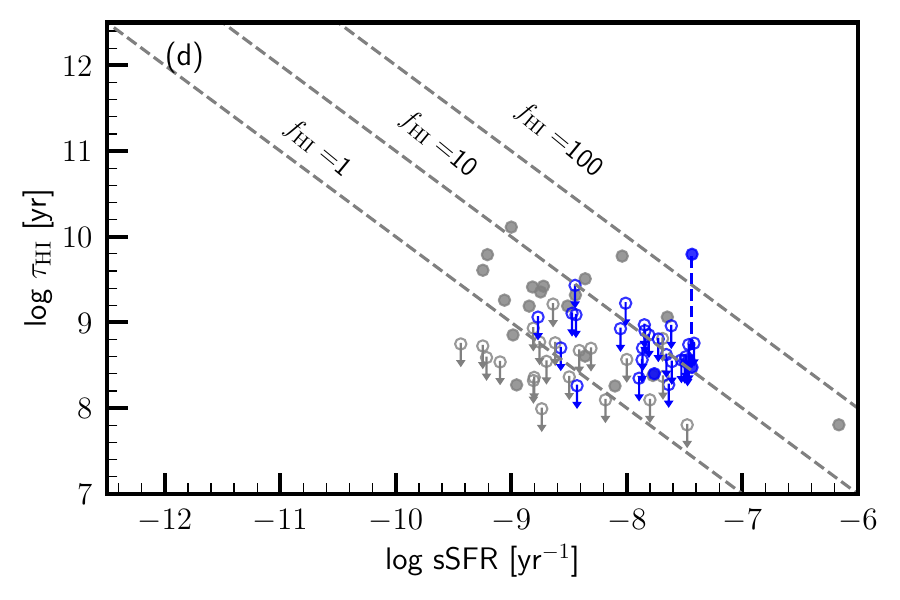}
    \caption{(a): $f_{\rm HI}$ vs. $M_{\ast}$. BBs observed with the FAST are shown with blue-colored circles. Gray-colored circles represent GPs with H{\sc i} study by \cite{kanekar2021ApJ...913L..15K}. Filled and empty circles represent detections and nondetections, respectively. Median values of log $f_{\rm HI}$ for respective median log $M_{\ast}$ values for main-sequence galaxies from \cite{catinella2018MNRAS.476..875C} are shown with black connected dots. Similarly, gray connected error bars mark the median log $f_{\rm HI}$ values at low and high stellar mass bins for GPs, respectively. The error bars represent the median absolute deviation (MAD) values of log $f_{\rm HI}$ at low and high stellar mass bins. For BB J1132+0809, we show the two $f_{\rm HI}$ values, one with the consideration that there is no neighboring galaxy (higher $f_{\rm HI}$) and the other in the case that the neighboring blue galaxy is at a similar redshift (lower $f_{\rm HI}$), connected with the dashed blue vertical line. (b):$f_{\rm HI}$ vs. sSFR. Symbols mean the same as in panel (a). Median values of log $f_{\rm HI}$ for respective median log sSFR values for main-sequence galaxies from \cite{catinella2018MNRAS.476..875C} are shown with black connected dots. Dashed gray diagonal lines mark the H{\sc i} depletion time scale ($\tau_{\rm HI}$). (c): $\tau_{\rm HI}$ vs. $M_{\ast}$. The dashed horizontal black line marks the median $\tau_{\rm HI}$ for main-sequence galaxies from \cite{saintonge2016MNRAS.462.1749S} and \cite{catinella2018MNRAS.476..875C}. Dashed horizontal magenta, cyan, and orange colored lines mark the median $\tau_{\rm HI}$ for mid-infrared bright BCDs in the Arecibo sample, all mid-infrared bright BCDs and other non-mid-infrared bright dwarf irregulars or BCDs, respectively, from \cite{chandola2024MNRAS.527..603C}. The gray dashed horizontal line marks the median $\tau_{\rm HI}$ value for GPs from \cite{kanekar2021ApJ...913L..15K}. For BB J1132+0809, we show two values of $\tau_{\rm HI}$, connected with the blue dashed vertical line for the two cases mentioned earlier. (d):$\tau_{\rm HI}$ vs. sSFR. Dashed gray diagonal lines mark the H{\sc i} gas fraction. Other symbols mean the same as in panel (c).}
    \label{fig:fhitauhi}
\end{figure*}

\section{Results and discussion}
\label{results}
Of the 28 sources observed with the FAST, we detected H{\sc i} emission with greater than 5$\sigma$ significance toward only two galaxies, namely J1026+0426 (BB 54 in \cite{yang2017ApJ...847...38Y}) and J1132+0809 (BB 68 in \cite{yang2017ApJ...847...38Y}). The profile toward J1026+0426 resembles a typical profile for a spiral galaxy and the profile toward J1132+0809 is an asymmetric profile. J1026+0426 has a neighboring spiral (Sbc) galaxy SDSS J102640.64+042716.2 of stellar mass\footnote{Estimated using the same method as \cite{chandola2024MNRAS.527..603C}.} $\sim$10$^{9.2}$ $M_{\odot}$ at an angular distance of $\sim$43.$\arcsec$6 (projected distance $\sim$36.3 kpc) from J1026+0426 and $z\sim$ 0.04222 \citep{ahumada2020ApJS..249....3A}.
 The 3$\arcmin$ beam of FAST cannot distinguish the H{\sc i} associated with J1026+0426 and the spiral galaxy. The optical systemic velocity of spiral galaxy $\sim$12657.2 km s$^{-1}$ is closer to the H{\sc i} velocity\footnote{Obtained as the average of velocities corresponding to the FWHM.} ($V_{\rm HI}$) $\sim$12662.3 km s$^{-1}$ obtained from the profile suggesting the possibility of H{\sc i} gas largely being associated with it (see Fig.~\ref{fig:hispectra1} upper panel). However, the peak flux density in the profile is close to the optical velocity of BB J1026+0426, suggesting that some H{\sc i} gas is also associated with the BB. For J1132+0809, we searched for neighboring galaxies within a 1.\arcmin5 \hspace{1mm} radius for possible confusing sources. In the SDSS optical image, there is a southern small blue galaxy (hereafter blue galaxy) J113259.13+080930.6 at an angular distance $\sim$12\arcsec\hspace{1mm}with no spectroscopic redshift but a photometric redshift of 0.004$\pm$0.0346 in SDSS DR16 \citep{ahumada2020ApJS..249....3A}. Since the photometric redshift of the blue galaxy has a large difference with the H{\sc i} velocity and optical spectroscopic redshift of BB J1132+0809, it suggests that a significant amount of H{\sc i} at the redshift of $\sim$0.04944 is associated with the BB J1132+0809. However, we also checked H{\sc i} in the FAST spectrum at the photometric redshift within the error range and did not detect any H{\sc i}. Hence, further optical spectroscopic observations would be required to verify the redshift.  If the photometric redshift is incorrect and assuming that the blue galaxy has a redshift corresponding to the H{\sc i} velocity, we estimate its stellar mass to be $\sim$10$^{8.5}$ $M_{\odot}$. We also have a tentative marginal 3$\sigma$ signal toward J1347+0755 (BB 88) which needs to be confirmed from deeper observations. We consider it as a nondetection for our analysis. The spectrum toward J0827+1059 is partly affected by bandpass ripples  
 and we consider it as a nondetection. J1509+3731 (BB 10) has been observed earlier by \cite{kanekar2021ApJ...913L..15K} and \cite{dutta2024MNRAS.531.5140D} where \cite{dutta2024MNRAS.531.5140D} has a detection from deeper GMRT observations. In our observations, the noise levels are higher for this source and hence we provide a nondetection spectrum for J1509+3731. For the detections, we provide the H{\sc i} parameters such as  H{\sc i} velocity ($V_{\rm HI}$), FWHM, integrated line flux density, and peak flux density in the notes of Table \ref{sourchar1}. We used the Arecibo IDL routine \texttt{mbmeasure} for estimating these parameters. We estimated the errors on these parameters using the method given in \cite{koribalski2004AJ....128...16K}.
 We estimate the H{\sc i} masses using the equation 
\begin{equation}
     M_\mathrm{HI} \sim 2.36 \times 10^{5}\,\times \left(\frac{D_{\rm L}}{\rm Mpc}\right)^{2} \times \left(\frac{F_{\rm HI}}{\rm Jy\,km\,s^{-1}}\right){M}_{\odot} , 
\label{eq1}
\end{equation}
where  
\begin{eqnarray}
F_{\rm HI}=\int \frac{S(v)}{\rm Jy}\frac{{\rm d}v}{\rm km\,s^{-1}}
\end{eqnarray}
  is the integrated line flux density  and $D_{\rm L}$ is the luminosity distance. The estimated H{\sc i} masses for J1026+0426 and J1132+0809 from the H{\sc i} profiles, are 6.9$\times$10$^{9}$ $M_{\odot}$ and 3.6$\times$10$^{9}$ $M_{\odot}$, respectively. We calculate the expected H{\sc i} mass of J1026+0426 to be $\sim$5.4$\times$10$^{7}$ $M_{\odot}$ assuming that the H{\sc i} mass ratio of the BB and its neighboring galaxy is equal to their stellar mass ratio. Similarly, in the case of J1132+0809, if the photometric redshift of neighboring blue galaxy is incorrect and has a redshift similar to the BB or H{\sc i} velocity, the expected H{\sc i} mass of BB would be $\sim$1.7$\times$10$^{8}$ $M_{\odot}$.
We have listed the spectral RMS noise ($\Delta S_\mathrm{rms}$)  per channel and H{\sc i} mass upper limits for all 26 BBs with nondetections in Table \ref{sourchar1}, and shown their spectra in Figure~\ref{fig:hispectra2}. The RMS noise varies from 0.15 mJy to 1.29 mJy per 10 km s$^{-1}$. The upper limits on H{\sc i} masses are estimated using the equation,
\begin{equation}
M_\mathrm{HI} \sim 2.36 \times 10^{5} \times \left(\frac{D_{\rm L}}{\rm Mpc}\right)^{2}\times 3\Delta S_\mathrm{rms}\Delta v \sqrt{\frac{50}{\Delta v}}    {M}_{\odot},
\label{eq2}
\end{equation}
where $\Delta v(=$ 10 km s$^{-1}$) is spectral velocity resolution in the spectra and we assume the typical width of the line to be 50 km s$^{-1}$. The median 3$\sigma$ upper limit on H{\sc i} mass is 2.0$\times$10$^{8}$ $M_{\odot}$.  We also list other properties such as stellar mass, SFR, H{\sc i} gas fraction, depletion time scale and O32 ratio obtained from \cite{yang2017ApJ...847...38Y} in Table~\ref{sourchar1}.  

\subsection{H{\sc i} gas fraction ($f_{\rm HI}$)}
The atomic hydrogen gas fraction or H{\sc i}-to-stellar mass ratio ($f_{\rm HI}= \frac{M_{\rm HI}}{M_{\ast}}$ ) has been used as a measure of fuel reservoir available per unit stellar mass and baryonic composition of galaxies in the literature.   It strongly depends on the stellar mass and star-forming nature of the galaxies \citep{chandola2024MNRAS.527..603C}. The galaxies with lower stellar mass are found to have higher H{\sc i} gas fraction implying atomic hydrogen is the dominant baryonic component \citep{saintonge2016MNRAS.462.1749S,thuan2016MNRAS.463.4268T,catinella2018MNRAS.476..875C,kanekar2021ApJ...913L..15K, chandola2024MNRAS.527..603C}. H{\sc i} gas fraction for main-sequence galaxies ($>$10$^{9}$ $M_{\odot}$) at nearby redshifts are found to be quite low, reaching up to $\sim$0.02 at the highest stellar mass ($\sim$10$^{11}$ $M_{\odot}$)  to $\sim$ 1 at $\sim$10$^{9}$ $M_{\odot}$ (see black dots connected with black line in Figure~\ref{fig:fhitauhi} (a)) in the extended Galaxy Evolution Explorer (GALEX) Arecibo SDSS Survey \citep[xGASS;][]{catinella2018MNRAS.476..875C}.  For the sample of GPs at the redshift $<$0.1, \cite{kanekar2021ApJ...913L..15K} found that gas fraction further increases at lower stellar mass Green Peas, consistent with the scenario of the dominance of atomic hydrogen in baryonic composition. For the BCDs of lower stellar masses ($\lesssim$10$^{8}$ M$_{\odot}$), the typical H{\sc i} gas fractions are found to be higher ($\sim$ 10) but with a lot of dispersion \citep{chandola2024MNRAS.527..603C,thuan2016MNRAS.463.4268T}.

We have plotted $f_{\rm HI}$ vs stellar mass and specific star formation rate (sSFR) in panels (a) and (b), respectively, of Figure~\ref{fig:fhitauhi}. In our study of BBs which have lower stellar masses than GPs, we made observations such that the galaxies following $f_{\rm HI}$-$M_{\ast}$ relation at lower stellar masses are detected with at least 3$\sigma$ significance. Hence, we found that the $f_{\rm HI}$ upper limits for BBs are similar to the upper limits for GPs for a similar common stellar mass range. The median 3$\sigma$ upper limit on $f_{\rm HI}$ for BBs is 10.2. It is lower than the expected value of $\sim$14.6 at the median stellar mass of 10$^{7.3}$ $M_{\odot}$ from extrapolating $f_{\rm HI}$-$M_{\ast}$ relation for GPs and main-sequence galaxies. 
Also, the majority of GPs and BBs due to their starburst nature appear to deviate from the $f_{\rm HI}$-sSFR relation for the main-sequence star-forming galaxies and have lower $f_{\rm HI}$ than expected. Considering the neighboring galaxy, we expect $f_{\rm HI}$ of 4.3 for J1026+0426. For J1123+0809, if there is no neighboring galaxy, we estimate $f_{\rm HI}$ to be $\sim$229. However, if the neighboring blue galaxy has a redshift corresponding to the H{\sc i} velocity, we expect a gas fraction of $\sim$11.0 for J1132+0809. We show both values of $f_{\rm HI}$ for J1132+0809 connected with vertical dashed lines in the upper panels of Figure~\ref{fig:fhitauhi}.

\subsection{H{\sc i} depletion time scale ($\tau_{\rm HI}$)}
H{\sc i} depletion time scale ($\tau_{\rm HI}=\frac{M_{\rm HI}}{\rm SFR}$) is a measure of how fast and efficiently the atomic gas reservoir is being used. 
In Figure~\ref{fig:fhitauhi} (lower panels), we have shown H{\sc i} depletion time scale vs stellar mass and sSFR. The expected H{\sc i} depletion time scale for J1026+0426 is 0.25 Gyr. For J1132+0809, the depletion time scale can be $\sim$6.3 Gyr if there is no neighboring galaxy, and $\sim$0.3 Gyr if the neighboring blue galaxy has a redshift corresponding to the H{\sc i} velocity. The median 3$\sigma$ upper limit on the depletion time scale from nondetections is 0.53 Gyr. This median upper limit on the H{\sc i} depletion timescale is similar to the depletion time scale of low redshift GPs by \cite{kanekar2021ApJ...913L..15K} and slightly higher than mid-IR bright BCDs in the Arecibo sample by \cite{chandola2024MNRAS.527..603C}. However, this is about 1 order of magnitude lower than $\tau_{\rm HI}$ for the local main sequence galaxies \citep{saintonge2016MNRAS.462.1749S,catinella2018MNRAS.476..875C} or other non-mid-IR bright dwarf irregulars or BCDs in the literature \citep{chandola2024MNRAS.527..603C}. This implies that the BBs have depleted their atomic gas reservoir or are in the process of depleting it quite fast.  
When the atomic gas depletion time scales ($\sim$5 Gyr) are larger than the molecular gas depletion time scales ($\sim$1 Gyr), as in the case of local main sequence galaxies, it has less effect on the star formation, as sufficient atomic gas is available for conversion to molecular gas \citep{saintonge2016MNRAS.462.1749S, catinella2018MNRAS.476..875C}. However, for starburst systems like BBs and GPs, shorter atomic gas depletion timescales are critical as the atomic gas might get converted faster into molecular gas due to favorable conditions like high density, and soon there could be a scarcity of atomic gas if not replenished from the nearby environment \citep{chandola2024MNRAS.527..603C}. It would require further observations of molecular gas toward these sources as the H{\sc i} to H$_{2}$ scaling relation for main-sequence galaxies may not be applicable in the case of extreme starburst, low metallicity BBs or GPs.
%%%%%%%%%%%%%%%
\begin{figure}
    \centering
    \includegraphics[scale=0.5]{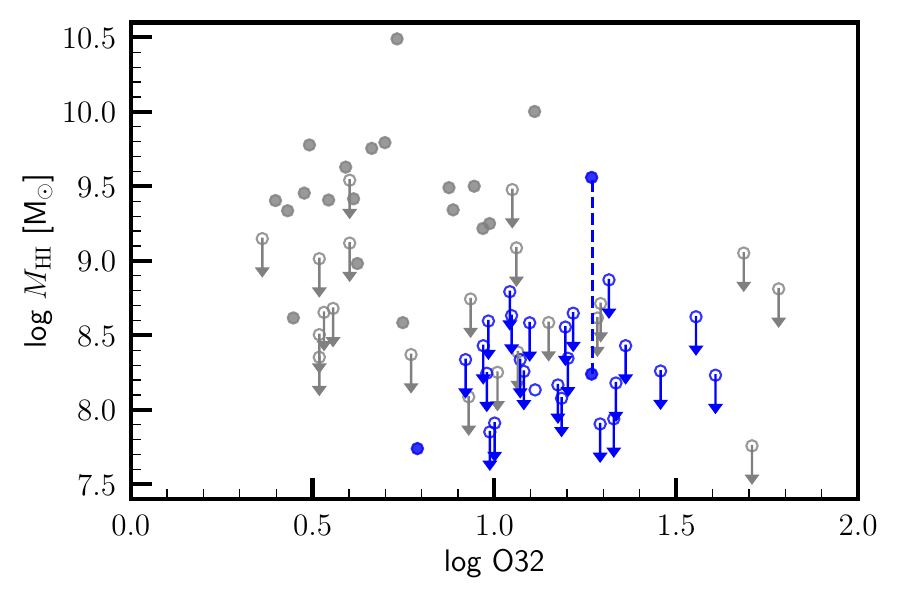}
    \includegraphics[scale=0.5]{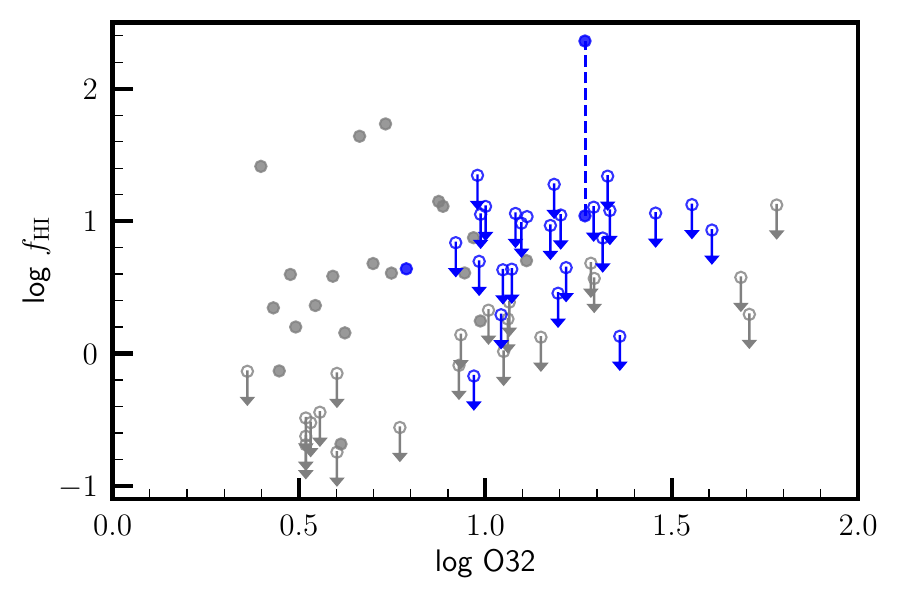}
    \includegraphics[scale=0.5]{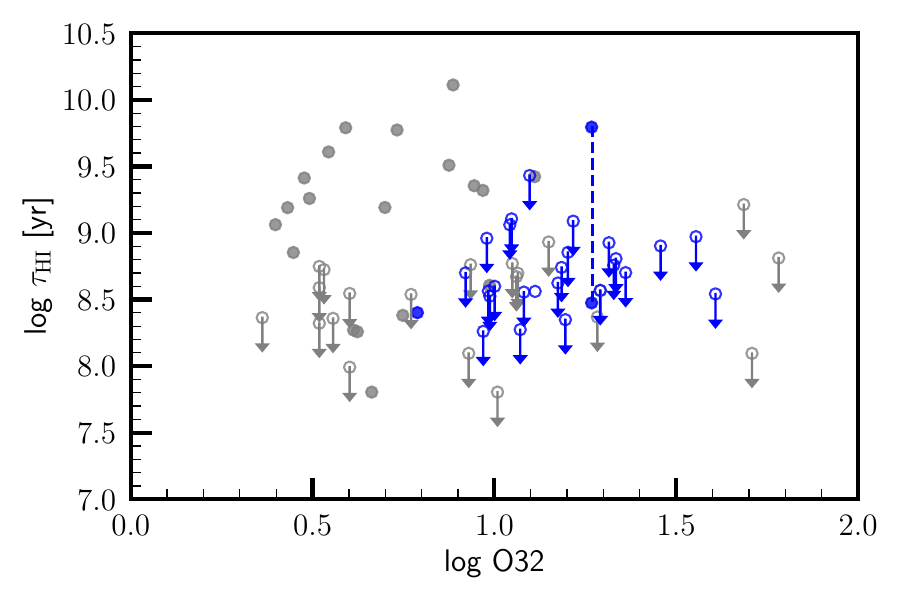}
    \caption{Top panel:$M_{\rm HI}$ vs. O32 $\equiv$ O[{\sc iii}]$\lambda$5007/O[{\sc ii}]$\lambda$3727 ratio.
    Middle panel:$f_{\rm HI}$ vs. O32 $\equiv$ O[{\sc iii}]$\lambda$5007/O[{\sc ii}]$\lambda$3727 ratio.
    Bottom panel: $\tau_{\rm HI}$ vs. O32 $\equiv$ O[{\sc iii}]$\lambda$5007/O[{\sc ii}]$\lambda$3727 ratio. The symbols mean the same as in Fig.~\ref{fig:fhitauhi}.}
    \label{fig:O32}
\end{figure}
%%%%%%%%%%% 
\subsection{H{\sc i} properties, O32 ratio, and Lyman Escape fraction}
The H{\sc i} detection rate\footnote{The 1$\sigma$ error on detection rate is estimated using \cite{gehrels1986ApJ...303..336G} small number statistics for Poisson distribution.} in BBs (7.1$^{+9.4}_{-4.6}$ \%) is significantly lower compared to the previous detection rate of $\sim$48\% for GPs by \cite{kanekar2021ApJ...913L..15K} when observed at similar levels of H{\sc i} mass and depletion time scale (or star-formation efficiency) upper limits. This detection rate is also low compared to low metallicity BCDs of similar stellar masses reported earlier in the literature \citep[e.g.][]{thuan2016MNRAS.463.4268T}. \cite{kanekar2021ApJ...913L..15K} reported that GPs with an O32 ratio higher than 10 have lower detection rate ($\sim$9\%) than those with lower O32 values ($\sim$62\%).   Of the 28 sources observed, we have O[{\sc ii}]$\lambda$3727 flux values available for only 26 sources from \cite{yang2017ApJ...847...38Y}. We find that all 26 BBs have an O32 ratio greater than 8.3,  and the majority (20/26) of the BBs have O32 values $\gtrsim$10. Hence, our finding of a low detection rate for BBs is similar to that for GPs of O32 $\gtrsim$10  by \cite{kanekar2021ApJ...913L..15K}. Of the two detections, J1026+0426 has an O32 value $\sim$6.2, while J1132+0809 has an O32 value $\sim$18.6, only the third H{\sc i} detection of BBs or GPs with an O32 ratio $\gtrsim$ 10 in the literature, the previous two being J1200+2719 by \cite{kanekar2021ApJ...913L..15K} and J1509+3731 by \cite{dutta2024MNRAS.531.5140D}. 

We have also plotted $M_{\rm HI}$, $f_{\rm HI}$ and $\tau_{\rm HI}$ vs O32 in the upper, middle, and lower panels of Figure~\ref{fig:O32}, respectively. The median 3$\sigma$ upper limit on $M_{\rm HI}$  for BBs with higher O32 is lower than $M_{\rm HI}$ values for GPs with lower O32. This is expected, as O32 tends to have higher values for lower stellar mass \citep{nakajima2014MNRAS.442..900N,Paalvast2018A&A...618A..40P,shen2024arXiv241023349S}, and high O32 selection for BBs gives lower stellar masses \citep{yang2017ApJ...847...38Y,izotov2016MNRAS.461.3683I}.
Hence, the lower $M_{\rm HI}$ values or upper limits for BBs are according to the $M_{\rm HI}$-$M_{\ast}$ relation \citep[e.g.][]{huang2012ApJ...756..113H}.  We further plot $M_{\rm HI}$ normalized with $M_{\ast}$ i.e. gas fraction ($f_{\rm HI}$) vs O32. In this plot, the upper limits on $f_{\rm HI}$ appear to increase with increasing O32 or decreasing stellar mass. Hence, the low H{\sc i} detection rates at higher O32 could be the effect of both $f_{\rm HI}$ sensitivity and ionization parameter indicator O32 \citep{kewley2002ApJS..142...35K,nakajima2014MNRAS.442..900N,shen2024arXiv241023349S}. In the common O32 range, the $f_{\rm HI}$ upper limits for BBs are slightly higher than GPs, suggesting the need for deeper observations and indicating $f_{\rm HI}$ lower than expected from $f_{\rm HI}$-$M_{\ast}$ as shown in Figure~\ref{fig:fhitauhi} panel (a). In the $\tau_{\rm HI}$ vs O32 plot, we find the median 3$\sigma$ upper limit on $\tau_{\rm HI}$ for BBs is lower than $\tau_{\rm HI}$ values for GPs of lower O32.

The O32 ratio has been discussed in some previous studies as an indirect indicator of the Lyman escape fraction or leakage \citep{jaskot2013ApJ...766...91J,izotov2016MNRAS.461.3683I, izotov2018MNRAS.474.4514I, izotov2018MNRAS.478.4851I}. However, O32 has a weak correlation with Lyman escape fraction \citep{izotov2018MNRAS.478.4851I, izotov2021MNRAS.503.1734I} and there have been suggestions to combine O32 ratios with other proxies such as SFR surface density,  Ly$\alpha$ line widths, and Ly$\alpha$ peak velocity separation \citep{izotov2021MNRAS.503.1734I,flury2022ApJ...930..126F, yang2017ApJ...844..171Y} to search for LyC leakers. Nevertheless, a high O32 ratio is one of the conditions fulfilled by BBs for a possible high Lyman escape fraction, making them the high potential candidates. The chances of LyC or Ly$\alpha$ photons escaping from a galaxy or getting absorbed also depend upon the distribution of neutral hydrogen gas in the interstellar medium (ISM). The porous nature, or low column density \citep{henry2015ApJ...809...19H}, or low covering fraction \citep{mckinney2019ApJ...874...52M} in the distribution of H{\sc i} gas in the ISM can allow LyC or Ly$\alpha$ photons to escape. Recently, in the H{\sc i} study of the nearest LyC emitting source, Haro 11, \cite{lereste2024MNRAS.528..757L} found  H{\sc i} distribution with offset from the LyC production site due to mergers and facilitating the escape of Ly$\alpha$ or LyC photons.  The interferometric H{\sc i} studies of GP J0213+0056 by \cite{purkayastha2022ApJ...933L..11P} and BB J1509+3731 by \cite{dutta2024MNRAS.531.5140D} also suggest a similar scenario of disturbed H{\sc i} morphology with offset from the peak starburst region due to mergers or interaction. To explore this in detail for our detections, we need interferometric observations of the H{\sc i} gas distribution, and also an accurate determination of the redshift of the neighboring galaxy of BB 1132+0809.

\section{Summary}
\label{sec:summary} 
From our study of 28 BBs with FAST, we report the H{\sc i} detection toward two BBs, J1026+0426 and J1132+0809. For BB J1026+0426 we estimate its H{\sc i} mass to be $\sim5.4\times10^7$ $M_\odot$ considering H{\sc i} contribution from its neighboring galaxy to the total H{\sc i} mass of $\sim$6.9$\times$10$^{9}$ $M_\odot$. For BB J1132+0809 the photometric redshift of the neighboring galaxy is significantly different. The H{\sc i} mass has been estimated to be $\sim3.6\times10^9$ $M_\odot$. However, if the neighboring galaxy has a redshift similar to the BB, the H{\sc i} mass is estimated to be $\sim1.7\times10^8$ $M_\odot$. 
From the nondetections, we have a median 3$\sigma$ upper limit of 2.0$\times$10$^{8}$ $M_{\odot}$ on H{\sc i} mass. H{\sc i} gas fractions for BBs tend to
have lower values than expected from $f_{\rm HI}$-sSFR or $f_{\rm HI}$-$M_{\ast}$ relations for local main-sequence galaxies. These galaxies have a median 3$\sigma$ upper limit on H{\sc i} depletion time scale $\sim$0.5 Gyr, nearly 1 order of magnitude lower compared to $\tau_{\rm HI}$ for local main-sequence galaxies. The H{\sc i} detection rate for these galaxies is also significantly low ($\sim$7\%) as compared to higher stellar mass GPs by \cite{kanekar2021ApJ...913L..15K} but similar to GPs with high O32$\equiv$O[{\sc iii}]$\lambda$5007/O[{\sc ii}]$\lambda$3727 ratios in their sample.

%% Also note that the acknowledgement environment does not support long amounts of text. If you have a lot of people and institutions to acknowledge, do not use this command. Instead, create a new \section{Acknowledgments}.
\begin{acknowledgments}
\section*{Acknowledgments}
We thank the anonymous reviewer for the constructive and helpful comments that helped to significantly improve the manuscript. Y.C. thanks the Center for Astronomical Mega-Science, Chinese Academy of Sciences, for the FAST distinguished young researcher fellowship (19-FAST-02). Y.C. also acknowledges the support from the National Natural Science Foundation of China (NSFC) under grant No. 12050410259 and the Ministry of Science and Technology (MOST) of China grant no. QNJ2021061003L. C.W.T. is supported by NSFC grant Nos. 11988101 and 12041302. Y.Z.M. acknowledges the support from the National Research Foundation of South Africa with grant Nos. 150580, 159044, CHN22111069370 and  ERC23040389081. Y.C. thanks IUCAA for its hospitality while finishing this work. This work has used the data from the Five-hundred-meter Aperture Spherical radio Telescope (FAST). FAST is a Chinese national mega-science facility, operated by the National Astronomical Observatories of the Chinese Academy of Sciences (NAOC). 

\end{acknowledgments}

\facility{FAST:500m.}
\software{Python, Numpy\citep{harris2020array}, Matplotlib\citep{Hunter:2007}, Astropy\citep{astropy:2013, astropy:2018, astropy:2022}, TOPCAT\citep{taylor2005ASPC..347...29T}}


%\bibliography{references}{}

\bibliographystyle{aasjournal}


%\appendix




\end{document}